\documentclass[sigconf,nonacm]{acmart}

\acmConference{}{}{}
\acmYear{}
\acmDOI{}
\acmISBN{}
\thanks{This is a preprint of a paper accepted for publication in the proceedings of the 2026 International Conference on the Social Impact of AI (SIAI 2026). This version does not include post-acceptance revisions or corrections and is not the final published version.}

\AtBeginDocument{%
  }

\usepackage{csquotes}
\usepackage{graphicx}
\usepackage{subcaption}
\usepackage{makecell}
\usepackage{booktabs} 
\usepackage{array}
\usepackage{caption}
\usepackage{tabularx}
\usepackage{pifont}

\newcolumntype{P}[1]{>{\centering\arraybackslash}p{#1}}

\begin{document}
\newcommand{\citeauthorNL}[1]{\NoHyper\citeauthor{#1}\endNoHyper}

\newcommand{\botName}[1]{AdvisorX}

\title{\botName{}: A Human-in-the-Loop AI-Powered Chatbot
to Support Academic Advising in Higher Education}


\author{Wendan Jiang}
\orcid{xxxx-xxxx-xxxx}
\affiliation{%
  \institution{Tufts University}
  \city{Medford}
  \state{MA}
  \country{United States}
}

\author{Shiyuan Wang}
\affiliation{%
  \institution{Tufts University}
  \city{Medford}
  \state{MA}
  \country{United States}
}

\author{Hiba Eltigani}
\affiliation{%
  \institution{Tufts University}
  \city{Medford}
  \state{MA}
  \country{United States}
}

\author{Rukhshan Haroon}
\affiliation{%
  \institution{Tufts University}
  \city{Medford}
  \state{MA}
  \country{United States}
}

\author{Abdullah Bin Faisal}
\affiliation{%
  \institution{Tufts University}
  \city{Medford}
  \state{MA}
  \country{United States}
}

\author{Fahad Dogar}
\affiliation{%
  \institution{Tufts University}
  \city{Medford}
  \state{MA}
  \country{United States}
}


\begin{abstract}
Academic advising is critical to student success, yet high student-to-advisor
ratios limit advisors' capacity to provide timely support. Prior work has applied
large language models (LLMs) to academic advising with promising results, showing
that careful design can produce personalized responses grounded in institutional
sources. However, existing AI-augmented advising systems are primarily
student-facing and focus on student experience, leaving advisors, a central
stakeholder in academic advising, out of both design and evaluation. To address
this gap, we introduce AdvisorX, an LLM-powered chatbot that automates
time-consuming advising tasks---such as information collection, policy lookup,
and response drafting---while keeping advisors in the loop to preserve human
oversight. We evaluated AdvisorX through a user study with eight advisors. Most
advisors rated the system as useful, and half reported that their initial
concerns eased after hands-on use. Our findings also show that accuracy alone
does not determine advisors' willingness to delegate: in situations involving
emotional distress or time pressure, such as a student facing academic failure,
advisors preferred to respond personally, valuing the relational dimension of
advising over response quality. We conclude with design implications for
advisor-facing advising tools and directions for future work.
\end{abstract}




\maketitle

\section{Introduction}
Academic advising is vital to student success and career development in higher
education, with positive advising experiences linked to improved academic
performance and stronger institutional integration~\cite{awadh2019,jamaludin2021,mu2016}.
Yet delivering high-quality advising remains challenging. Many institutions lack
professional advisors, shifting the burden to faculty who often have limited
training in institutional policies~\cite{hartbaldridge2020,he2017}. In addition,
high student-to-advisor ratios contribute to emotional exhaustion and diminished
personal accomplishment among advisors~\cite{soria2024,zhang2023dual}.

Given the increasing use of generative AI in a wide range of settings, it is
natural to consider whether it can be used for academic advising. While early
studies show the promise of generative AI in educational
scenarios~\cite{chiang2024,henkel2024llm,kazemitabaar2023}, academic advising
presents a more challenging setting due to two key factors. First, existing
generative AI models~\cite{anthropic2024claude3,achiam2023gpt4} are not grounded in
authoritative institutional knowledge, making them prone to hallucinations in
institution-specific queries~\cite{akiba2023ai,joren2025sufficient}. Inaccuracies in academic
advising can have serious consequences, such as delaying graduation or derailing
career paths. Second, effective advising requires personalization based on each
student's unique circumstances~\cite{personalization-jones}. General-purpose language
models, without access to student-specific information, may produce factually
correct but generic responses that fail to address individual needs. Existing
AI-augmented advising systems have shown promising results on both challenges. To
reduce hallucination, several systems~\cite{advisely2025,Nittanyadvisor,argobot2025}
use retrieval-augmented generation (RAG)~\cite{RAG} to ground their
responses in institutional documents. To personalize responses, some
systems~\cite{aaai-recommender2025,seahawk2024} also draw on each student's academic record, such as
completed coursework and degree progress, so that answers reflect the student's
own situation rather than general policy.

Yet grounding and personalization address only the informational side of
advising. Advising research has long treated advising as relational rather than
purely informational, with students' connection to their advisor central to its
effectiveness~\cite{crookston2009,drake2011}. We therefore argue that AI is
better positioned to assist advisors than to replace them, taking on the
time-consuming work of retrieving information and drafting responses while
advisors review and approve each draft before it reaches a student. However, most
existing AI-augmented advising systems are student-facing: they leave advisors, a
central stakeholder in academic advising, out of the response workflow. Their
evaluations likewise focus on student experience, with little attention to
advisors' perspectives. How advisors perceive such a system, and when they are
willing to rely on it, remains an open question.

In this paper, we investigate these questions through AdvisorX, an LLM-powered
advising chatbot that grounds its responses in authoritative institutional
sources, personalizes them using student profiles maintained across
conversations, and routes every draft to an advisor for review before it reaches
the student. We conducted an exploratory study with eight advisors (four staff,
four faculty) in graduate computer science advising at a private U.S.\
university. The study addresses three research questions:
\begin{itemize}
  \item \textbf{RQ1}: How do advisors perceive the integration of generative AI
    into academic advising, and how do these perceptions change after hands-on use?
  \item \textbf{RQ2}: What factors shape advisors' comfort with letting AI handle
    student queries?
  \item \textbf{RQ3}: What design implications for AI-assisted advising emerge
    from advisors' experiences with AdvisorX?
\end{itemize}

Our findings suggest that although advisors initially approached AI with caution,
four of eight reported that their concerns eased after hands-on use, and five of
eight rated AdvisorX as useful (RQ1). However, advisors' willingness to let
AdvisorX handle a query depended on more than response accuracy (RQ2). For a
query from a student at risk of failing a course, most advisors judged the draft
accurate but still preferred to respond personally, citing the complexity and
urgency of the situation and the student's need for human engagement. Advisors'
experiences also point to two design implications (RQ3). First, systems should
decline to answer when sources are insufficient. Advisors read hedged answers as
misleading, not cautious. Second, systems should let advisors adjust the length
and tone of drafts to align with their advising styles. More broadly, AI should
support advisors' work, not replace it. Advising depends on relationships, and
those are built through advisors' direct connection with students.
\section{Related Work}

This work draws on research on academic advising and on AI systems designed to
support it.

\subsection{Academic Advising in CS Education}

Academic advising is a critical part of higher education, supporting students in
navigating their educational program in a manner that helps them achieve their
goals. It serves as a first line of defense in majors like computer science,
where students face a higher risk of dropping out~\cite{mendez2023impressions,schmidt2025ms}.
Because of resource constraints, institutions often cannot hire enough full-time
advisors~\cite{akiba2023ai}, and much of the advising load falls to faculty, who
must balance it against growing student populations, research expectations, and
teaching duties~\cite{camp2017generation,zhang2023dual}. In interviews with engineering faculty advisors, Zhang~\cite{zhang2023dual} found that advisors struggle to provide
personalized guidance within limited time, often leading to rushed meetings that
fail to address students' complex needs.

Beyond workload pressures, advising also demands substantial information
retrieval and synthesis. Advisors must keep track of a wide range of resources
and requirements to provide accurate planning, which may require additional
training whenever new requirements are introduced~\cite{brodley2022,zhang2023dual}.
M\'endez et al.~\cite{mendez2023impressions} observed that advisors must synthesize
information about students' academic histories, personal circumstances, and
career aspirations while considering program requirements, prerequisite chains,
and course availability. These challenges suggest a role for tools that
consolidate distributed resources and handle queries requiring synthesis across
them---work that is time-consuming and error-prone when done by hand.

\subsection{Relational Dynamics of Advising}

Beyond its informational demands, advising is also relational. Advising research
has long distinguished prescriptive advising, in which the advisor diagnoses a
problem and supplies an answer, from developmental advising, a collaborative
relationship built on shared responsibility and trust~\cite{crookston2009}.
Crookston notes that direct answers suit specific informational questions, but
that the problem a student presents is often only a symptom of a deeper
one~\cite{crookston2009}. Students' relationships with their advisors have also
been linked to retention and persistence~\cite{drake2011}. Tools that handle
information retrieval and synthesis could therefore give advisors more time for
relationship-building, but only if they support advisors rather than stand in for
them.

\begin{figure*}[t]
    \centering
    \includegraphics[width=1\textwidth]{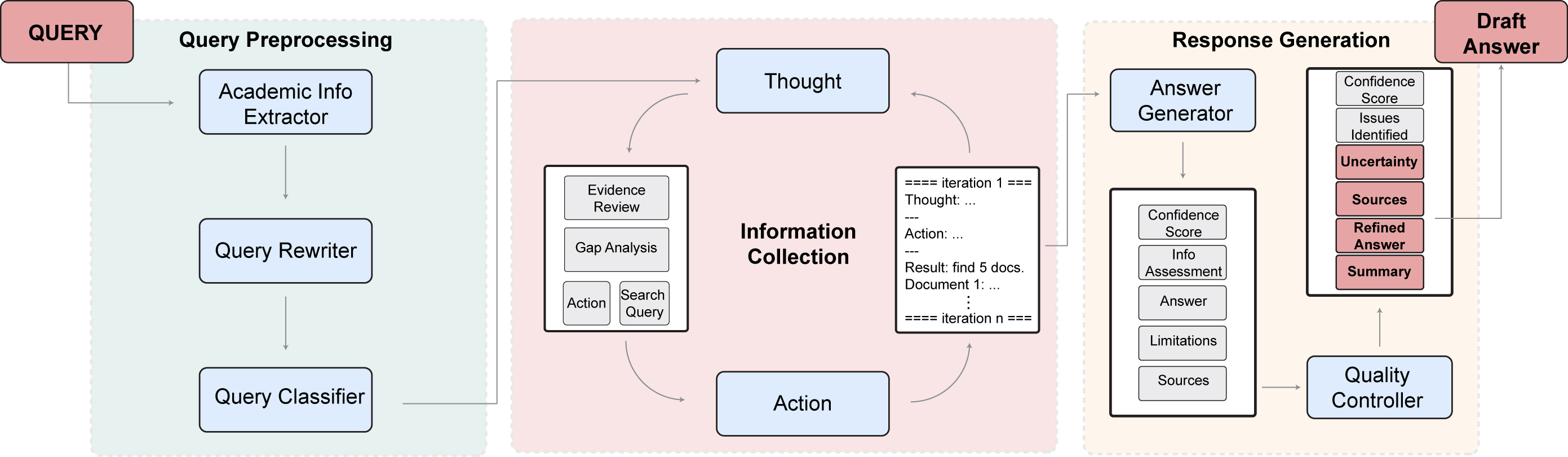}
    \caption{Three-phase multi-agent architecture. Queries classified as off-topic
    during Phase~1 bypass Phases~2--3. Red-highlighted components in the draft
    response (detailed answer, summary, cited sources, and uncertainty) are delivered
    to advisors for review.}
    \label{fig:architecture}
    \Description{A flowchart depicting a three-phase multi-agent architecture. A query enters Phase 1, Query Preprocessing (blue background), passing through three components: Academic Info Extractor, Query Rewriter, and Query Classifier. In Phase 2, Information Collection (red background), an iterative loop cycles between a Thought and an Action agent. In Phase 3, Response Generation (yellow background), an Answer Generator produces an output containing Confidence Score, Info Assessment, Answer, Limitations, and Sources, which is then reviewed by a Quality Controller. The final Draft Answer includes Confidence Score, Issues Identified, Uncertainty (red-highlighted), Sources (red-highlighted), Refined Answer (red-highlighted), and Summary (red-highlighted). Red-highlighted components are delivered to human advisors for review.}
\end{figure*}

\subsection{AI-Augmented Systems in Academic Advising}

Early explorations applied general-purpose LLMs directly to advising questions.
Akiba and Fraboni~\cite{akiba2023ai} typed frequently asked questions from a teacher
education program into the free version of ChatGPT and evaluated the answers.
ChatGPT performed well on broad questions that do not depend on a particular
institution or student, such as career outlook. Many advising questions, however,
hinge on a specific institution's policies, which a general-purpose model has no
access to. Subsequent systems close this gap by grounding responses in
institutional knowledge, most commonly through retrieval-augmented generation
(RAG)~\cite{RAG}. ARGObot~\cite{argobot2025} retrieves from the
university student handbook, falling back on web search when the handbook does
not cover a query, and Nittany Advisor~\cite{Nittanyadvisor} retrieves from a vector
database of university course data and academic websites.
Advisely~\cite{advisely2025} similarly pairs GPT-4o with an
institution-specific knowledge base of policies, rules, and curriculum details.

\newcommand{\cmark}{\ding{51}}
\newcommand{\xmark}{\ding{55}}

\begin{table}[t]
\centering
\caption{Comparison of AI-augmented advising systems.}
\label{tab:comparison}
\small
\setlength{\tabcolsep}{3pt}
\begin{tabular}{lcccc}
\toprule
\textbf{Study} &
\textbf{\makecell{Domain\\Knowledge}} &
\textbf{\makecell{Personal-\\ized}} &
\textbf{\makecell{Advisor-in-\\Loop}} &
\textbf{\makecell{Advisor\\Eval.}} \\
\midrule
Akiba and Fraboni~\cite{akiba2023ai}  & \xmark & \xmark & \xmark & \xmark \\
ARGObot~\cite{argobot2025}            & \cmark & \xmark & \cmark & \xmark \\
Advisor SeaHawk~\cite{seahawk2024}    & \cmark & \cmark & \xmark & \xmark \\
Advisely~\cite{advisely2025}          & \cmark & \xmark & \xmark & \xmark \\
Ma et al.~\cite{aaai-recommender2025} & \cmark & \cmark & \xmark & \xmark \\
Nittany Advisor~\cite{Nittanyadvisor} & \cmark & \xmark & \xmark & \xmark \\
Lekan and Pardos~\cite{lekan2025ai}   & \cmark & \cmark & \xmark & \cmark \\
\textbf{AdvisorX (Ours)}              & \cmark & \cmark & \cmark & \cmark \\
\bottomrule
\end{tabular}
\end{table}

Grounding tailors answers to an institution, but not yet to an individual
student. Lekan and Pardos~\cite{lekan2025ai} examined whether GPT-4 could
personalize major recommendations for undeclared students. They surveyed students
about their interests, course preferences, and career goals, and included each
student's responses in the prompt. Advisors rated the recommendations favorably,
and GPT-4 matched the advisor's choice for a third of students. However, some
advisors found that the responses lacked nuance and emphasized that good advising
depends on two-way dialogue with students, suggesting that a one-time survey
captures only part of what personalization requires. Other systems go beyond
one-time surveys and personalize responses more interactively. Advisor
SeaHawk~\cite{seahawk2024} lets students upload their academic records and then ask
follow-up questions, verifying prerequisites and tracking completed credits,
while Ma et al.~\cite{aaai-recommender2025} generate degree plans from a student's goals,
interests, and academic progress, which students can accept, modify, or refine
through feedback.

In existing systems, advisors play a peripheral role. Only
ARGObot~\cite{argobot2025} involves them through escalation: when a student
asks to speak with a human, it emails an advisor to schedule an appointment.
Advisors are similarly absent from evaluation, as these systems have been
assessed primarily on response quality and student experience
(Table~\ref{tab:comparison}). Only Lekan and Pardos~\cite{lekan2025ai} collected
perspectives from advisor side, and it was through a one-time survey.
Understanding how advisors experience such a system in practice, and when they
are willing to rely on it, requires going beyond one-time survey to an in-depth
user study in which advisors use the system hands-on.
\section{\botName{}}
\begin{figure*}[ht]
    \centering
    \includegraphics[width=1\textwidth]{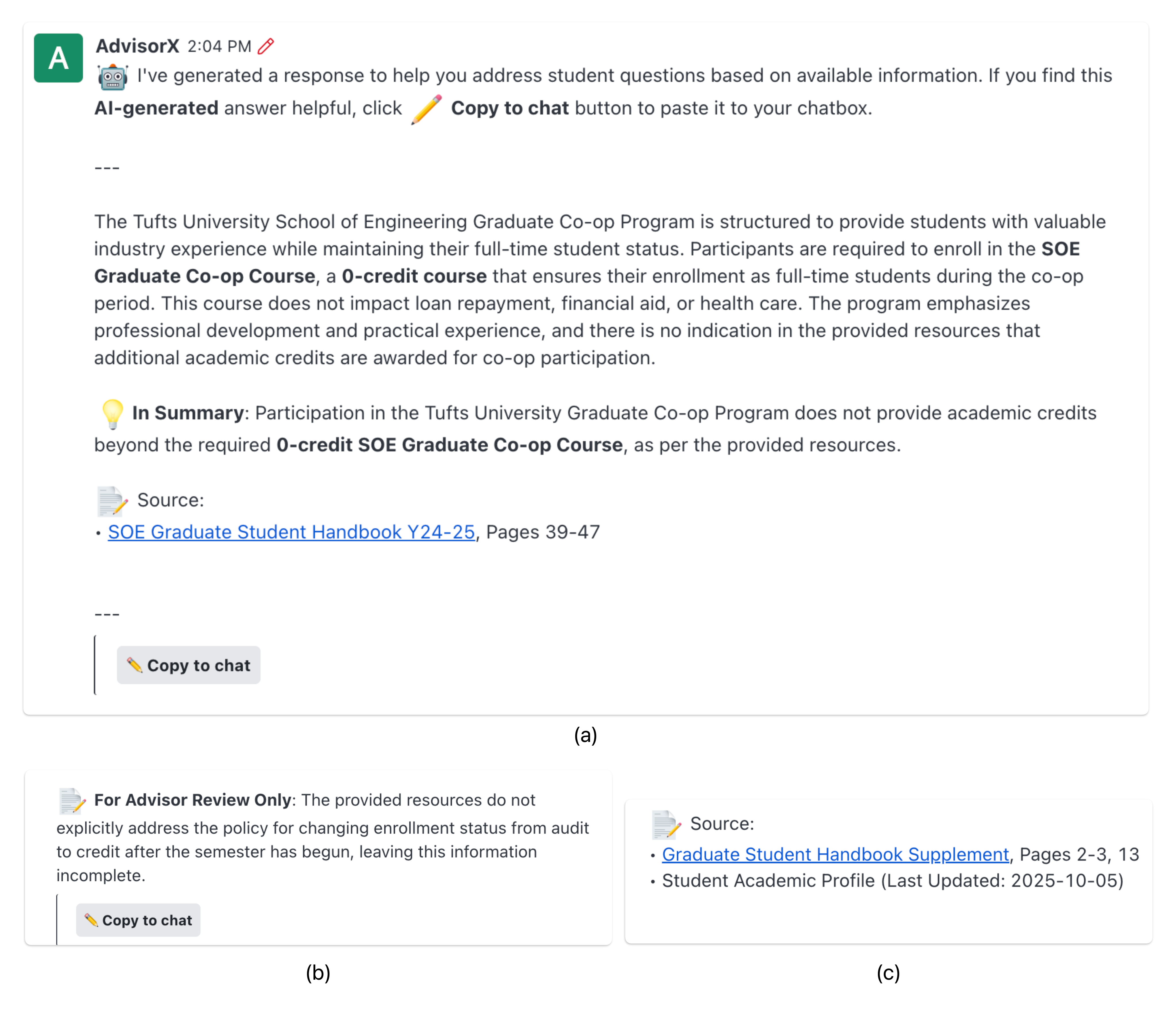}
    \caption{Advisor draft-review interface showing (a)~a complete draft response
    with cited sources, (b)~a note flagged ``For Advisor Review Only'' indicating
    missing information, and (c)~a timestamped reference to the student's academic
    profile.}
    \label{fig:interface}
\end{figure*}

AdvisorX connects students and advisors through Rocket.Chat~\cite{rocketchat},
an open-source messaging platform. When a student sends a message, the backend
processes it through three phases (Fig.~\ref{fig:architecture}).

\textbf{Query Preprocessing} prepares an incoming message for information
retrieval. Three agents handle distinct tasks. An \emph{Academic Info Extractor}
pulls academic details from the message to update the student profile, a
persistent record of each student's background (e.g., degree program) that is
incrementally built from conversations and stored in the backend database. A
\emph{Query Rewriter} makes the query self-contained, since messages in an
ongoing conversation often refer to earlier turns. A \emph{Query Classifier}
filters out off-topic questions so that advisors are not asked to review drafts
for messages that fall outside advising altogether. \textbf{Information
Collection} gathers what the answer requires. A ReAct~\cite{yao2023react} loop
iterates between a \emph{Thought Agent} and an \emph{Action Agent}: the Thought
Agent uses chain-of-thought reasoning~\cite{wei2022}, informed by the student
profile, to decide what information is still needed. The Action Agent then
selects among four actions: retrieving from an institutional knowledge base using
RAG, querying a structured course database, searching the web when institutional
sources do not cover the topic, and asking the student directly when
student-specific details are missing. In \textbf{Response Generation}, an
\emph{Answer Generator} synthesizes the collected information and the student
profile into a detailed draft, followed by a \emph{Quality Controller} that
checks for hallucinations and proper citations. The system uses
GPT-4o-mini~\cite{achiam2023gpt4} together with GPT-4o to balance cost and accuracy.

The resulting draft is sent to the designated advisor. As shown in
Fig.~\ref{fig:interface}a, each draft includes four components, each supporting a
different part of the review task: (1)~a detailed response with key information
highlighted, (2)~a one-line summary that lets an advisor triage a draft without
reading it in full, (3)~a list of cited sources with document names, page
numbers, or web links, so that claims can be checked against the original, and
(4)~an optional note flagging missing information when the system cannot fully
answer the question, directing the advisor's attention to the parts that warrant
scrutiny (Fig.~\ref{fig:interface}b). When a draft response draws on a student's
stored academic profile, a timestamped reference is displayed so advisors can
verify that the information is up to date (Fig.~\ref{fig:interface}c). Advisors
can then review, edit, or approve the draft before sharing it with the student
via a ``Copy to Chat'' button.

\section{User Study}
We evaluated AdvisorX through a user study with advisors. All procedures of our
user study were approved by our university's Institutional Review Board (IRB). We
conducted semi-structured interviews (approximately 60 minutes each) with eight
advisors (four staff advisors, SA1--SA4, and four faculty advisors, FA1--FA4)
from the university's CS department. Staff advisors specialized in
university-wide policies, while faculty advisors focused on career guidance and
course recommendations. All were actively serving in advising roles.

\begin{figure*}[t]
\centering
\includegraphics[width=\textwidth]{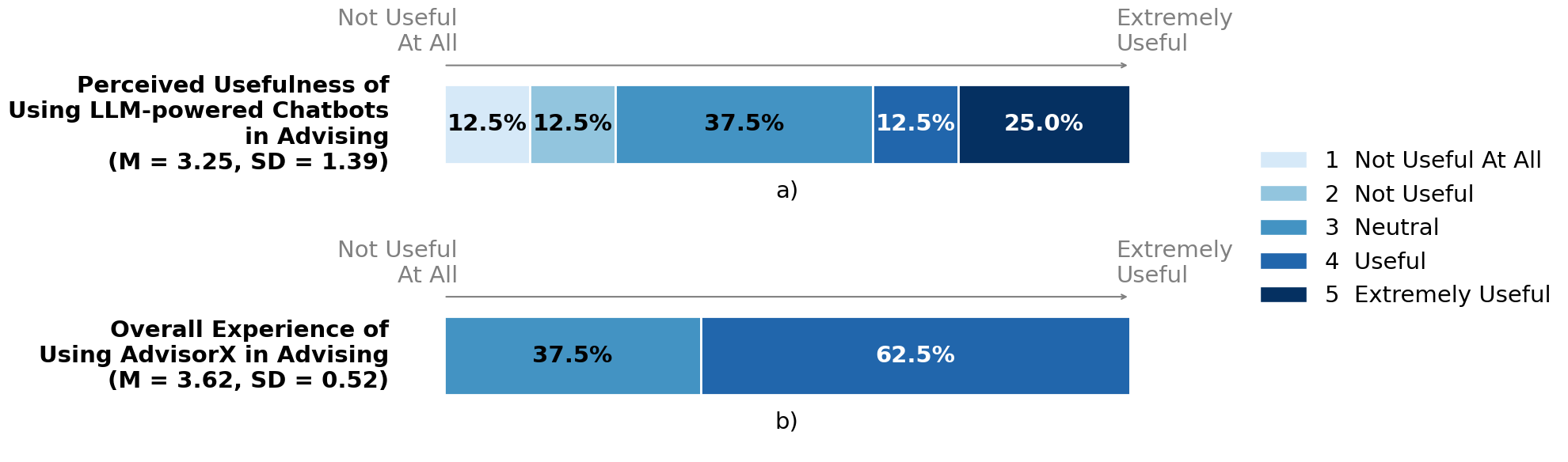}
\caption{Participants' ratings of (a) the perceived usefulness of LLM-powered chatbots in advising
and (b) their overall experience of using AdvisorX in advising, on a 5-point scale from 1 (Not
Useful At All) to 5 (Extremely Useful). Percentages indicate the share of participants selecting
each rating.}
\label{fig:ratings}
\end{figure*}

\subsection{Semi-structured Interviews}

The semi-structured interview consisted of three parts, with audio and screen
interactions recorded throughout. In Part~1, we captured participants' baseline
attitudes toward AI, including a rating of GenAI's perceived usefulness (5-point
Likert scale). In Part~2, participants were introduced to AdvisorX and then used
it to respond to four pre-selected advising questions, identical for all
participants. We varied the questions by information availability to examine how
advisors review drafts backed by different levels of evidence. A response
supported by a direct source may pose a different review problem than one
assembled from partial information or generated with no source at all. The four
questions were Q1 (Direct-Answer), where the knowledge base contains information
that directly answers the question; Q2 (Indirect-Answer), where related
information must be interpreted and synthesized; Q3 (No-Answer), where no
relevant information exists; and Q4 (Student-Specific), where answering requires
the student's own information (e.g., courses taken, declared major). For each
question, participants reviewed and optionally edited the system-generated
response, then completed a survey rating its accuracy, appropriateness of tone
and style, and their willingness to delegate (all on 5-point Likert scales),
followed by a discussion of the reasoning behind each rating. After all four
questions, participants provided an overall usefulness rating of AdvisorX. In
Part~3, we collected reflections on the prototype and ideas for AI integration in
advising.

\subsection{Data Analysis}

Audio recordings were transcribed and analyzed using inductive thematic
analysis~\cite{clarkbraun}. Following a similar approach~\cite{park2025}, two
members of the research team first reviewed two randomly selected interview
transcripts (25\% of participants) to develop an initial understanding of
participants' perspectives through open coding. Each coder independently
performed line-by-line analysis of one transcript to generate initial thematic
codes, after which they met to compare observations and collaboratively develop a
shared coding framework. Both researchers then re-analyzed the initial
transcripts using the framework to ensure consistency. The remaining transcripts
were divided evenly, with each researcher coding half, and the researchers met
again to discuss their codes. All themes were generated inductively from the data
without predetermined categories.

Quantitative data, including Likert ratings, were analyzed with descriptive
statistics (means and standard deviations) to identify where ratings diverged
across question types; we did not conduct significance testing given the sample
size.
\section{Results}
We report findings from eight advisors, four faculty advisors (FA1--FA4) and four
staff advisors (SA1--SA4). We first describe advisors' perceptions of generative
AI in advising: the concerns they brought to the study and how these shifted
after using AdvisorX (RQ1). We then examine how advisors evaluated the generated
drafts in terms of perceived accuracy and style, and finally what shaped their
willingness to delegate different types of questions (RQ2).

\subsection{Advisor Perceptions of GenAI in Advising (RQ1)}

\begin{figure*}[h]
    \centering
    \includegraphics[width=1\textwidth]{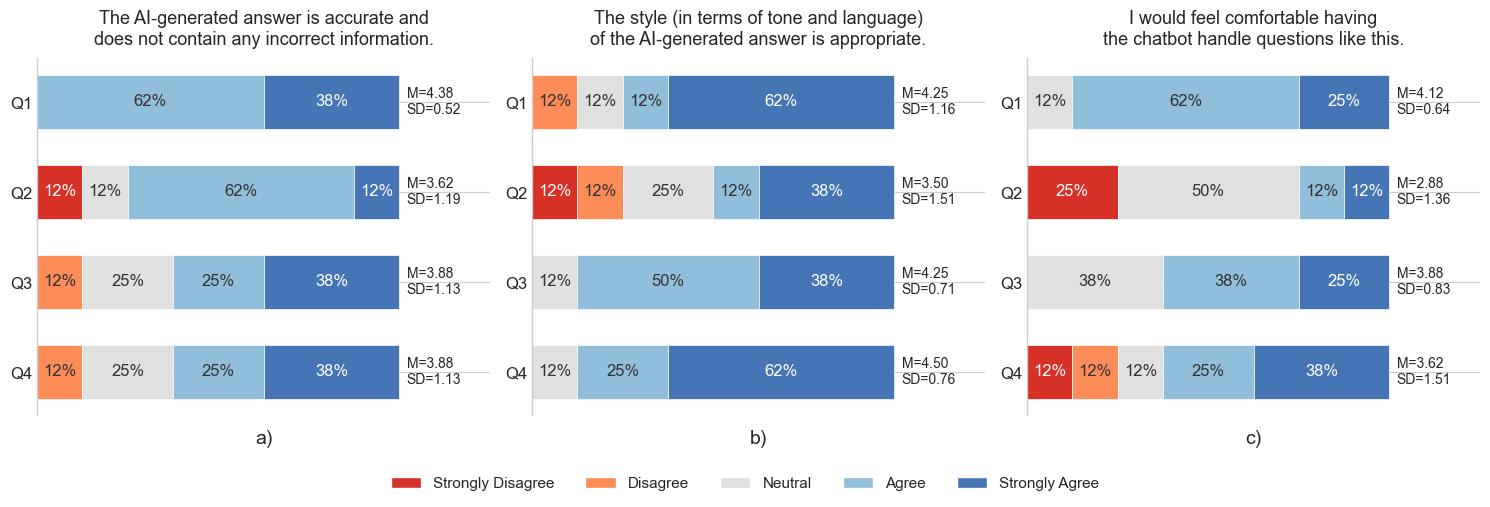}
    \caption{Participants' ratings of (a)~the perceived usefulness of LLM-powered
    chatbots in advising and (b)~their overall experience of using AdvisorX in
    advising, on a 5-point scale from 1 (Not Useful At All) to 5 (Extremely Useful).
    Percentages indicate the share of participants selecting each rating.}
\label{fig:usefulness}
    \Description{Three stacked horizontal bar charts, each with four bars labeled Q1 through Q4, using a five-point Likert scale colored from Strongly Disagree (red), Disagree (orange), Neutral (gray), Agree (light blue), to Strongly Agree (dark blue). Panel (a) is titled "The AI-generated answer is accurate and does not contain any incorrect information," with medians of 4.38, 3.62, 3.88, and 3.88 for Q1 through Q4. Panel (b) is titled "The style (in terms of tone and language) of the AI-generated answer is appropriate," with medians of 4.25, 3.50, 4.25, and 4.50 for Q1 through Q4. Panel (c) is titled "I would feel comfortable having the chatbot handle questions like this," with medians of 4.12, 2.88, 3.88, and 3.62 for Q1 through Q4. Each bar segment is labeled with its percentage value, and median (M) and standard deviation (SD) are shown alongside each bar.}
    \label{fig:boxplots}
\end{figure*}

\paragraph{Prior Experience and Concerns of Using GenAI}
Most participants (7 of 8) reported prior experience with generative AI chatbots.
As shown in Fig.~\ref{fig:usefulness}a, initial attitudes toward GenAI's
usefulness in advising were mixed ($M = 3.25$, $SD = 1.39$). Five of eight
advisors rated its usefulness at or below neutral, indicating a generally
cautious stance. The two who assigned the maximum rating (5, Extremely Useful)
did so conditionally: FA2 considered AI useful only if deployed directly to
students with a disclaimer about potential inaccuracies, while SA3 saw value only
for Direct-Answer questions.

A central source of skepticism was the accuracy of AI-generated responses. As FA2
emphasized, ``I don't care if the bot is like accurate 99\% of the time, because
that 1\% is going to make it completely useless for me.'' Advisors worried that
correcting errors could ultimately increase rather than reduce their workload.

Advisors also noted that advising often involves deeply personal questions such
as ``what do I want to do as a career?'' (FA4). For these individualized
situations, they worried that AI cannot reliably ``figure out what directions
you're heading and give you good advice'' (FA3), given that ``no two students are
the same'' (SA3). They concluded that generic advice is unlikely to provide
substantial value.

\paragraph{Shifts in Advisor Attitudes}
Half of the advisors explicitly noted that their prior concerns had been eased
after using AdvisorX. We identified two broad patterns in how attitudes shifted.
First, advisors who had initially expressed skepticism about AI accuracy reported
more positive opinions, finding the system ``does a better job of pulling up the
factual'' (FA1) and ``more accurate than I was expecting'' (SA1). Clear citations
and the system's transparency---particularly its acknowledgment of
uncertainty---helped ease their concerns. Second, advisors concerned about
GenAI's ability to handle nuanced, personalized situations found their concerns
``slightly eased'' (SA3), partly because the master's program is ``a little more
standard'' (SA4). They appreciated the system's proactive requests for student
information and found its personalization for graduate-level advising ``pretty
solid'' (SA3). However, concerns were only partially alleviated, as undergraduate
advising is ``a bit more personalized than just what classes have you taken''
(SA3), raising doubts about whether AdvisorX would generalize to that context.

Advisors responded positively to AdvisorX overall: 5 of 8 rated the system's
usefulness as 4 on a 5-point scale, with the remaining three rating it as neutral
(Fig.~\ref{fig:usefulness}b). No advisor gave a negative rating. Several advisors
expressed strong enthusiasm, describing the system as having ``exceeded my
expectations'' (SA3) and as ``something that could truly help'' (SA4). However,
hands-on use did not resolve all of advisors' concerns. Advisors voiced three
major concerns after using the system, some carried over from before the study
and others that arose during use. First, regarding reliability: FA4 argued that
hallucinations are ``fundamental'' to LLMs and could not be fully resolved
through system design without improvements to the underlying models. Second,
regarding efficiency: some advisors questioned whether the system truly reduces
workload, noting that if every response requires verification, ``it didn't help
me at all'' (FA3). Third, regarding the advising relationship: advisors
emphasized that AI tools in general should not diminish the interpersonal
dimension of advising. As FA4 noted, advising sessions are sometimes ``the only
mechanism we have to get our undergrads to talk to us'', and removing that touch
point could have negative consequences.

\subsection{How Advisors Evaluated Generated Drafts}

\paragraph{Perceived Accuracy}
The user study captured perceived accuracy, as faculty advisors were less
familiar with detailed university policies. As shown in
Fig.~\ref{fig:ratings}a, Q1 (Direct-Answer) received the highest and most
consistent accuracy ratings ($M = 4.38$, $SD = 0.52$), with all advisors judging
the responses as accurate. Ratings for the remaining three question types were
lower and more variable. Two issues emerged. First, some advisors rated responses
as ``incorrect information'' (SA1) when they contained hedging language (e.g.,
``may''), despite the system explicitly flagging such uncertainty in ``For
Advisor Review Only'' notes. Second, when no relevant information was available
(Q3), advisors preferred the system to decline answering altogether rather than
attempt a response with flagged uncertainty, fearing that students might
``interpret [\ldots] incorrectly'' (FA2).

\paragraph{Style and Tone}
Style and tone ratings were shaped by both advisor role and question context.
Regarding advisor role, faculty and staff advisors held different preferences for
response detail. Faculty advisors (FA2, FA3) tended to perceive AI-generated
responses as overly verbose across all four question types, while staff advisors
(SA3, SA4) more often preferred detailed responses, arguing that providing
information beyond the literal wording of a student's question can ``answer
numerous questions that the student may have'' (SA4) and that ``the student
probably needs that information and doesn't realize it'' (SA3).

Regarding question context, these role-based preferences converged when the
system handled questions where information was unavailable (Q3). In Q3, most
advisors---including SA3, who generally favored detailed responses---explicitly
preferred concise output, explaining that additional content ``feels less
useful'' (SA3) when no definitive answer was available. Question context also
influenced perceptions of emotional tone. As shown in Fig.~\ref{fig:ratings}b,
Q2, which concerned a scenario involving potentially failing a course, exhibited
the widest spread in style and tone ratings ($M = 3.50$, $SD = 1.51$). Half of
the advisors reported that the tone was inadequate. SA2 noted that the tone felt
too ``flat'' and suggested creating ``more urgency,'' while FA4 emphasized that
such scenarios should prioritize student well-being---understanding ``what
problems did they run into during the semester''---before presenting procedural
rules.

\subsection{What Shapes Willingness to Delegate (RQ2)}

Delegation willingness varied across question types, with advisors' concerns
clustering around two factors: the urgency of the situation and the reliability
of information sources. As shown in Fig.~\ref{fig:ratings}c, Q1 (Direct-Answer)
received the highest delegation ratings ($M = 4.12$, $SD = 0.64$). Q2
(Indirect-Answer) received the lowest ratings ($M = 2.88$, $SD = 1.36$), driven
by the complexity and seriousness of the scenario, which involved potentially
failing a course. As FA1 explained, when a response requires ``multiple bullet
points,'' students ``should definitely talk to an advisor at that point, because
it already shows how complicated the situation is.'' Q3 (No-Answer) received
somewhat positive ratings ($M = 3.88$, $SD = 0.83$). Although advisors preferred
outright refusal from an accuracy standpoint, the system's explicit
acknowledgment of uncertainty nonetheless made delegation feel safer---even for
AI-cautious advisors like SA1. Q4 (Student-Specific) showed the widest spread
($M = 3.62$, $SD = 1.51$). While most advisors were willing to delegate, some
worried that responses relied too heavily on student-provided information, which
may be incomplete or inaccurate.
\section{Discussion}
This study examined AI-augmented advising from the perspective of the advisors
who review its output. In this section, we first discuss the design implications
of these findings (RQ3), then reflect on the role of human connection in advising
and the cost of oversight.

\subsection{Design Implications (RQ3)}

Advisors' experiences with AdvisorX point to three design implications for
AI-assisted advising. First, advisor-facing interfaces should adapt to individual
advising styles. Advisors differed in their preferred response length, level of
detail, and tone, so adjustable verbosity and tone settings would let each
advisor tailor drafts to their own practice. Beyond individual preferences, tone
should also adapt to situational context, as emotionally sensitive scenarios
(e.g., academic failure) may require more supportive responses than routine
factual inquiries. Lastly, when information is unavailable, less is more.
Although the system flagged its uncertainty rather than withholding a response,
advisors preferred it to clearly acknowledge its limitations and stop. Hedging
language (e.g., ``may'') was similarly counterproductive: advisors often
interpreted it as unhelpful or misleading, rather than as appropriate caution.

\subsection{The Irreplaceable Role of Human Connection}

Our findings suggest that in emotionally sensitive or urgent situations such as
academic failure, advisors preferred to respond to students personally, not
primarily because they doubted the system's accuracy, but because they saw these
cases as requiring human engagement, such as emotional support, that AI cannot
provide. This aligns with prior work in nursing education showing that students'
acceptance of AI-generated help depends on more than response
quality~\cite{lambert2026}. Our findings extend this observation to academic
advising and to the perspective of advisors: even when advisors judged a draft
accurate, their willingness to delegate depended on whether the situation called
for a human connection.

Advisors also valued advising sessions as a way to stay connected with students;
as FA4 noted, these sessions are sometimes the only way to get undergraduates to
talk to their advisors. This echoes the long-standing view of advising as
relational rather than purely informational~\cite{crookston2009,drake2011}. Taken
together, these results suggest that the human-relational dimension of
advising---not just correctness---is a critical factor of successful integration
of generative AI in advising.

\subsection{The Cost of Oversight}

Several advisors noted that reviewing and verifying AI-generated responses can be
cognitively demanding. An advisor-in-the-loop design that requires review of
every query may not make effective use of advisors' attention: research on
automation shows that sustained monitoring of a reliable system erodes the very
scrutiny such oversight depends on~\cite{Raja2010complacency}. Our delegation findings
suggest that review effort could be allocated more selectively. Although
advisors' willingness to delegate varied across queries, they converged on
Direct-Answer queries, which are straightforward and backed by direct sources.
Such queries also make up a substantial share of common advising questions: Shah
et al.~\cite{Nittanyadvisor} classified 40\% of frequently asked advising questions as
information-based and answerable from institutional resources. Review therefore
need not be applied uniformly. A system could instead route queries by type,
answering routine informational requests without advisor review while reserving
advisor attention for cases in which advisors themselves prefer to intervene.

\section{Limitations and Future Work}

This exploratory study intentionally prioritized depth over scale to gain rich
qualitative insight into advisors' attitudes, which resulted in a small sample of
eight advisors from a single institution. Advisors also reviewed drafts for
predefined scenarios rather than real student queries. Future work could address
these limitations through longitudinal deployments with real student queries and
studies across additional institutions and advising contexts.

\section{Conclusion}

In this paper, we bring advisors into the design and evaluation of AI-augmented
advising, a stakeholder previously absent from both. Through AdvisorX, an
advisor-in-the-loop advising assistant, and a study with the advisors who used
it, we examined how advisors perceive such systems, what shapes their willingness
to delegate, and what their experiences imply for design. We found that hands-on
use eased some of advisors' concerns, but that their willingness to delegate
depended on more than accuracy: in urgent or emotionally sensitive situations,
advisors preferred to respond personally. These findings suggest that AI can best
support advising by handling routine information work, leaving advisors free to
focus on the relationships that advising depends on.

\begin{acks}
We thank Tony Li for his helpful discussions. We also thank the students in
the Spring 2025 GenAI for Social Impact course at Tufts University for their
comments. This work was partially funded by NSF Award 2106797.
\end{acks}


\bibliographystyle{ACM-Reference-Format}
\bibliography{main}


\end{document}